\documentclass[letter, structabstract]{aa} 
\usepackage[varg]{txfonts}
\usepackage{graphicx}
\bibpunct{(}{)}{;}{a}{}{,} 
\usepackage{amssymb}

\newcommand{\herschel}{\textit{Herschel}}

\newcommand{\textsim}{$\sim$}
\newcommand{\kms}{km~s$^{-1}$}

\newcommand{\tlambda}{$\lambda$} 
\newcommand{\farcmRA}{$\hbox{$.\!\!^{\mathrm{s}}$}$} 
\newcommand{\arcm}{^{\prime}}			
\newcommand{\eu}{$E_\mathrm{u}$} 		
\newcommand{\dlq}[1]{``}
\newcommand{\drq}[1]{''}
\newcommand{\isowater}{{H$_2^{18}$O}}
\newcommand{\normalwater}{H$_2$O}
\newcommand{\dimethylether}{CH$_3$OCH$_3$}

\newcommand{\trans}[6]{$#1_{#2,#3}-#4_{#5,#6}$}
\begin{document}
   \title{Warm water deuterium fractionation in IRAS~16293-2422}
\subtitle{The high-resolution ALMA and SMA view}
   \author{M.~V. Persson
          \inst{1}\fnmsep\inst{2}
          \and
          J.~K. J{\o}rgensen\inst{2}\fnmsep\inst{1}
          \and E.~F. van Dishoeck
          \inst{3}\fnmsep\inst{4}
          }

   \institute{Centre for Star and Planet Formation,
   Natural History Museum of Denmark,
   University of Copenhagen,
   {\O}ster Voldgade 5-7, \\DK-1350,
  Copenhagen K, Denmark\\
              \email{magnusp@nbi.dk}
         \and
            Niels Bohr Institute, University of Copenhagen, Juliane
            Maries Vej 30, DK-2100 Copenhagen {\O}, Denmark
         \and
             Leiden Observatory, Leiden University, P.O. Box 9513, NL-2300 RA Leiden, The Netherlands
        \and
            Max-Planck Institute f\"{u}r extraterrestrische Physik (MPE), Giessenbachstrasse, 85748 Garching, Germany
             }

   \date{Received October 29, 2012; accepted November 28, 2012}

 
  \abstract
   {Measuring the water deuterium fractionation in the inner warm regions of low-mass protostars has so far been hampered by poor angular resolution obtainable with single-dish ground- and space-based telescopes. Observations of water isotopologues using (sub)millimeter wavelength interferometers have the potential to shed light on this matter.}
   {To measure the water deuterium fractionation in the warm gas of the deeply-embedded protostellar binary IRAS~16293-2422. }
   {Observations toward IRAS~16293-2422 of the
     \trans{5}{3}{2}{4}{4}{1} transition of \isowater\ at
     692.07914~GHz from Atacama Large Millimeter/submillimeter Array
     (ALMA) as well as the \trans{3}{1}{3}{2}{2}{0} of \isowater\ at
     203.40752~GHz and the \trans{3}{1}{2}{2}{2}{1} transition of HDO
     at 225.89672~GHz from the Submillimeter Array (SMA) are
     presented. 
     }
   {The 692~GHz \isowater\ line is seen toward both components of the
     binary protostar. Toward one of the components, \dlq{}source~B\drq{}, the line is seen in absorption toward the continuum,
     slightly red-shifted from the systemic velocity, whereas emission
     is seen off-source at the systemic velocity. Toward the other
     component, \dlq{}source~A\drq{}, the two HDO and \isowater\ lines
     are detected as well with the SMA. From the \isowater\
     transitions the excitation temperature is estimated at
     $124\pm12$~K. The calculated HDO/\normalwater\ ratio is
     ($9.2\pm2.6)\times10^{-4}$ -- significantly lower than
     previous estimates in the warm gas close to the source. It is
     also lower by a factor of $\sim$5 than the ratio deduced in the
     outer envelope.}
   {Our observations reveal the physical and chemical structure of
     water vapor close to the protostars on solar-system scales. The
     red-shifted absorption detected toward source B is indicative of
     infall. The excitation temperature is consistent with the picture
     of water ice evaporation close to the protostar. The low HDO/H$_2$O ratio deduced here suggests that the differences between the inner regions of the protostars and the Earth's oceans and comets are smaller than previously thought.}

   \keywords{astrochemistry -- stars: formation -- protoplanetary disks -- ISM: abundances -- ISM: general
 }
   \maketitle
%

\section{Introduction}

Water plays an essential role for life as we know it, but its origin
on Earth is still unclear: was water accreted during the early stages
of Earth\rq{s} formation, or brought by smaller solar system bodies
such as comets at later times? To deduce the origin of Earth's water
and the amount of chemical processing it has experienced, one option
is to measure the water deuterium fractionation (HDO/H$_2$O) during
different stages in the evolution of protostars and compare it to what
we measure in Earth\rq{}s oceans and comets.

Generally the HDO/H$_2$O ratio in Earth's oceans of $3\times 10^{-4}$
\citep[e.g.,][]{lecuyer98} and Oort cloud comets of 8.2$\times
10^{-4}$ \citep{villanueva09} are found to be enhanced above the D/H\footnote{A HDO/H$_2$O ratio of \drq{}$x$\dlq{} corresponds to a D/H ratio of \dlq{}$x/2$\drq{}.}
ratio in the protosolar nebula $\sim 1.5\times 10^{-5}$
\citep{linsky03,geiss98} due to deuterium fractionation processes. The
factor of 2 higher abundance ratio in the Oort cloud comets than in
Earth's oceans has previously been taken as an indication that only a
small fraction of Earth\rq{}s water was delivered by comets. Recently
however, a HDO/\normalwater\ ratio of $3.2\times 10^{-4}$ was measured
for the Jupiter class comet Hartley~2 with the \herschel\ Space
Observatory \citep{hartogh11} and $4.2\times 10^{-4}$ for the Oort
cloud comet Garradd \citep{bockelee12}, indicating values closer to
those of Earth\rq{}s water. 

Attempts at measuring the water deuterium fractionation in protostars
have resulted in different conclusions. \citet{parise03} used
ground-based infrared observations of the stretching bands of OH and
OD in water ice in the outer parts of envelopes and found \emph{upper
  limits} ranging from 0.5\% to 2\% for the HDO/H$_2$O ratios in four
embedded low-mass protostars. In the gas-phase it is possible to
detect lines of HDO, but such studies differ on the interpretation
with HDO/H$_2$O ratios in protostars ranging from cometary values
\citep{stark04}, to a few \% \citep{parise05,liu11}. Even more
recently, \citet{coutens12} deduced a HDO/H$_2$O ratio in
IRAS~16293-2422 of $3.4\times 10^{-2}$ in the inner parts and
$0.5\times 10^{-2}$ in the outer envelope by modeling a large range of lines observed with \herschel.

One problem with previous measurements of HDO/\normalwater\ is the
relatively large beam size of single-dish ground- and space-based
telescopes. Spherically symmetric power-law models of protostellar
envelopes have usually been employed to interpret the observations.
While such models are appropriate to interpret continuum and line
emission on larger scales (\textgreater300~AU), they are not suited to
unambiguously analyze the observed compact components since there are
clear indications that they are not an accurate representation of the
conditions on small scales
\citep[e.g.,][]{jorgensen05,chiang12}. Estimates of abundance ratios
on these smaller scales are thus subject to significant uncertainties
due to extrapolations of the underlying physical structures.

High angular resolution millimeter wavelength aperture synthesis
observations offer a possibility to circumvent this issue. Recently
\citet{jorgensen10a} detected the water isotopologue \isowater\ toward
the deeply embedded protostar NGC-1333~IRAS4B on scales of \textless
50~AU using the IRAM Plateau de Bure Interferometer (PdBI), which
combined with an upper limit on the HDO column density from the SMA resulted in a 3$\sigma$ upper limit to the
HDO/H$_2$O abundance ratio of 6$\times 10^{-4}$. To follow-up these
results we initiated an extended survey of the H$_2^{18}$O and HDO
emission on arcsecond scales using the IRAM PdBI and SMA
\citep{persson12a}.

IRAS~16293-2422 is a Class~0 protostellar binary (sep $\sim5\arcsec$, 600~AU) located 120~pc away in the LDN~1689N cloud in the $\rho$~Ophiucus star-forming region \citep{knude98,loinard08}. With a rich spectrum at (sub)millimeter wavelengths \citep{blake94,vandishoeck95,cazaux03,chandler05,caux11,jorgensen11} it has been one of the prime targets for studies of astrochemistry during the star-formation process, revealing the presence of a range of complex organic species \citep{bottinelli04,kuan04,bisschop08} and prebiotic molecules \citep{jorgensen12} on (sub)arcsecond scales.

In this letter we present, for the first time, \emph{high-resolution}
ground-based observations of several isotopologues of water toward
IRAS~16293-2422 with both the ALMA (\textsim0\farcs2; 24~AU) and the SMA
(\textsim2\farcs3; 276~AU). We derive direct, model-independent
estimates of the water excitation temperature and use it to calculate
the column density and the HDO/\normalwater\ ratio in the warm inner
envelope.


\section{Observations}

Observations of the deeply-embedded low-mass protostellar binary IRAS~16293-2422 were carried out at 690~GHz with ALMA and 230~GHz with the SMA, targeting the \trans{5}{3}{2}{4}{4}{1} (692.07914~GHz) and \trans{3}{1}{3}{2}{2}{0} (203.40752~GHz) transitions of \isowater\ and the \trans{3}{1}{2}{2}{2}{1} (225.89672~GHz) transition of HDO (see Table~\ref{table:observations} in the appendix/electronic version). In the tables, the 203~GHz observations are indicated with \dlq{}1\drq{}, 225~GHz observations with \dlq{}2\drq{} and 692~GHz with \dlq{}3\drq{} (see Table~\ref{table:lines} and Table~\ref{table:observations})

The ALMA observations of IRAS~16293-2422 were conducted as part of the ALMA Science Verification (SV) program: IRAS~16293-2422 was observed with 13 antennas on April~16 and 17, 2012 in a seven pointing mosaic centered at $\alpha = 16^\mathrm{h} 32^\mathrm{m} 22\farcmRA7 $, $\delta = -24\degr28\arcm32\farcs5$ (J2000). The observations cover projected baselines from 26 to 402~m (62 to 945~k\tlambda). One of the basebands was centered at 691.299~GHz with a bandwidth of 1.875~GHz and a spectral resolution of 0.923~MHz (0.4~\kms), a setup that covers the \isowater\ \trans{5}{3}{2}{4}{4}{1}\ line at 692.07914~GHz. Calibration observations include the quasars 1924-292 and 3c279 for the bandpass, the asteroid Juno for the amplitude, and the quasars 1625-254 and nrao530 for the phase. The science verification data are available as calibrated {\it{uv}}-data sets, which were used in our analysis. The calibrated data were imaged using the CASA software package \citep{mcmullin07}.

The lower-lying excited \isowater\ \trans{3}{1}{3}{2}{2}{0} transition at 203.4075~GHz was observed with the SMA on May 1, 2011 in the compact configuration with seven antennas. This configuration resulted in projected baselines between 9 and 69 meters (6 to 47~k\tlambda). The passband was calibrated by observations of the quasar 3c279 while the absolute flux and complex gains were calibrated by observing Titan and the quasars 3c279, 1517-243 and 1626-298. The raw data calibration followed the standard recipes using the MIR package \citep{qi08} and then MIRIAD \citep{sault95} was used to subtract the continuum from the data to create continuum-free line maps. 

Finally, we utilized SMA observations of the HDO \trans{3}{1}{2}{2}{2}{1} line at 225.89672~GHz from the SMA \citep{jorgensen11}: those observations have a spectral resolution of 0.41~MHz (0.54~\kms) and cover the projected baselines between $8.6-119.6$~m ($6.5-90$~k\tlambda). For further information about those observations we refer to \citet{jorgensen11}. The resulting beam size, field of view, velocity resolution and RMS of the various line data are summarized in Table~\ref{table:observations} in the appendix/electronic version.

\section{Results}

\begin{figure}[ht]
   \centering
    \includegraphics[width=0.8\linewidth]{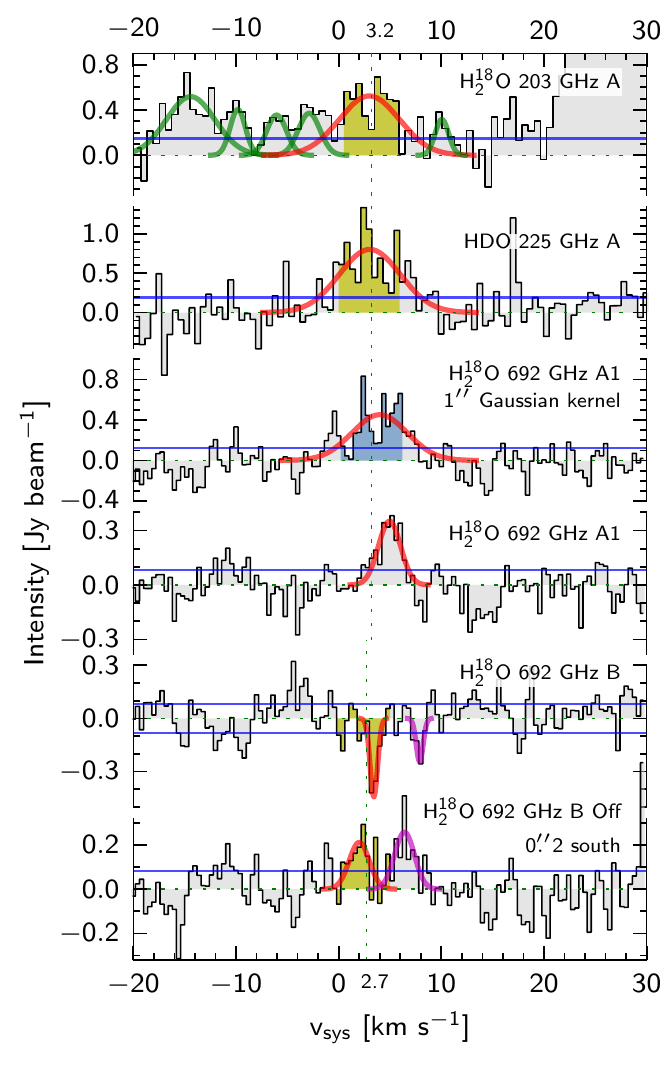}
      \caption{Spectra of the targeted water lines toward both IRAS~16293-2422 A and B. The 203~GHz \isowater\ line spectrum (top) was binned to twice the resolution for clarity. The 1D Gaussian fits (red: \isowater; green: CH$_3$OCH$_3$; magenta: Unidentified) and the RMS (blue) are plotted. Shaded areas show the interval over which the integrated intensities have been calculated (yellow and blue). The dotted green vertical lines shows $v_\mathrm{LSR}=3.2$ and $2.7$~\kms\ for source A and B respectively. The third spectrum from the top (blue fill) is from data smoothed with a 1\arcsec\ Gaussian kernel.}
         \label{figure:spectra}
   \end{figure}

Figure~\ref{figure:spectra} shows the spectra around the water lines toward the continuum peaks with Gaussian fits. Source A is in itself a binary \citep{chandler05} and we here refer to the components as A1 (the Northeast component) and A2 (the Southwest component) and the lower resolution single component as A. Both \isowater\ lines are clearly detected toward source A in emission, while only the 692~GHz \isowater\ line is seen toward source B in absorption. The \isowater\ emission is clearly associated with peak A1 in the 692~GHz data. The \isowater\ 692~GHz absorption line toward the continuum peak of source B is narrow and marginally red-shifted; the absorption is not due to the broad outflow seen in some lines \citep{jorgensen11}.

This weak absorption feature is not seen in the SMA data but the high sensitivity of ALMA combined with the stronger continuum at high frequencies is enough to detect it. The lines were identified using the JPL and CDMS catalogs through the Splatalogue compilation \citep{pickett98,muller01}.

   The Splatalogue compilation shows no other possible line candidates
   at least $\pm1$~\kms\ from the line peak positions of the
   \isowater\ line at 203~GHz and the HDO line at 225~GHz. Lines that
   fall within $\pm2-3$~\kms\ exist, but are of too low intrinsic line
   strengths or lack additional components that should have been
   detected at other velocity offsets e.g., (CH$_3$)$_2$CO,
   $^{13}$CH$_3$CH$_2$CN or C$_6$H. At 203.403 and 203.410 GHz (i.e.,
   about 10 and -3~\kms) lies \dimethylether\
   \trans{3}{3}{1}{2}{2}{1}EA/EE, which is accounted for when fitting the 203~GHz line. These lines are narrow enough to avoid any
   significant interference with \isowater\ (see top most plot in
   Fig.~\ref{figure:spectra}).

   For the \isowater\ ALMA data at 692~GHz the line identification in
   source A is complicated by the fact that the source is resolved and
   shows a systematic velocity pattern in other lines
   \citep{pineda12}. The line toward the continuum position of source
   A is red-shifted by about 2~\kms\ from the systemic $v_{LSR}$ of
   3.2~\kms, similar to lines from other species (see
   Fig.~\ref{figure:spectra}, panel four). The third spectrum from the
   top in Fig.~\ref{figure:spectra} shows the spectrum after smoothing
   with a Gaussian kernel of 1\arcsec\ size. The Gaussian fit to these
   data peaks closer to $v_{LSR}$=3.2~\kms, indicative indeed of
   resolved velocity structure. Outflow emission \citep[c.f.,][]{loinard12}, as also seen in H$_2$O masers on small scales \citep{alves12}, could explain this velocity shift, but it is not possible to deduce its importance from the current data.

   However, another possibility is that line blending causes the shift. Toward source B, the off position spectrum shows two emission lines, the second line at $\sim6$~\kms\ is unidentified (bottom spectrum Fig.\ref{figure:spectra}). The \isowater\ line at 692 GHz toward source A may therefore also be affected by line blending. However, the unambiguous detection of the 692~GHz line toward source B and the SMA observations of water toward source A suggest that at least half of the 692 GHz feature toward  source A is indeed due to the \isowater\ line.

\begin{figure}[ht]
   \centering
    \includegraphics[width=0.9\linewidth]{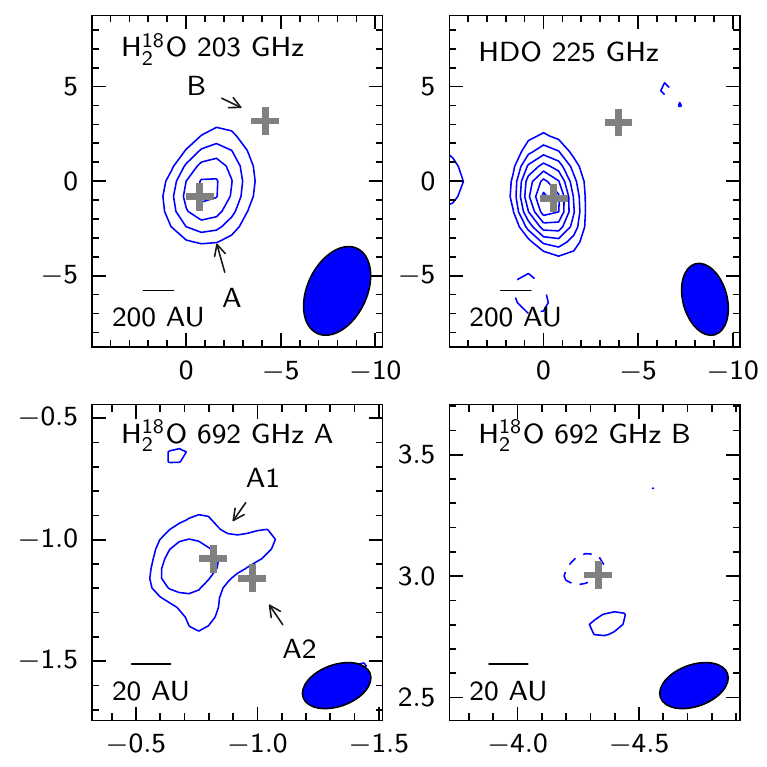}
      \caption{Integrated intensity maps for all observed water lines calculated from channels $\pm3$ from the systemic velocity, deduced from previous observations \citep{jorgensen11}. The absorption line toward source B was integrated in the interval $0-5$~\kms. Note the different spatial scales of the top vs the bottom panels. The beam is shown in the lower right corner and the gray crosses show the position of  continuum peaks from elliptical Gaussian fits. Units on the axes are offset in arcseconds from the phase center of the 203~GHz observations. Contours are in steps of $2\sigma$ starting at $3\sigma$, dashed contours represent negative values. 
              }
         \label{figure:moments}
   \end{figure}

Table~\ref{table:lines} lists parameters from Gaussian fits to the image and {\sl uv}-plane of the line emission. The integrated intensity maps for the detected water lines are shown in Fig.~\ref{figure:moments}. 

   \begin{table}[ht]
	\caption[]{Parameters from fits to the integrated maps and spectra of the different water lines for source A and B. The given errors are the statistical uncertainties and the 20\% calibration uncertainty in intensity is not included. The spectral line fits were done toward the continuum peak and integrated intensities deduced from circular Gaussian fits in the \textit{uv}-plane (203 and 225~GHz) or an elliptical Gaussian fit in image plane (692~GHz).}
         \label{table:lines}
         \begin{center}
         \begin{tabular}{lllll}
            \hline
            \noalign{\smallskip}
            Line		&  Size 						& Intensity 		&  Line width				 \\
            Id  		& {[\arcsec]} 					& {[Jy \kms]} 		& {[\kms]}		 \\
            \noalign{\smallskip}
            \hline

            \noalign{\smallskip}
            1 (A)		& $1.7\pm0.7$   & \ \ $3.9\pm1.4$	    & $6.9\pm1.5$					\\
            2 (A)		& $1.1\pm0.4$   & \ \ $8.9\pm1.8$  	& $7.0\pm0.7$				 \\
            3	 (A1)	& $0.2\times0.4\, (13\degr)$ & \ \ $3.4\pm0.6$      & $6.3\pm0.7$				\\	
            3	 (B)	& Point\ fit     & $-0.43\pm0.07$ 	& $0.9\pm0.2$					\\

            \noalign{\smallskip}
            \hline
         \end{tabular}
         \end{center}
\tablefoot{The column \dlq{}Line~Id\drq{} gives the source indicated in parentheses, and the number corresponds to one of the lines (see text and Table~\ref{table:observations}). The line width is the FWHM from a Gaussian fit to the spectral line. The elliptical Gaussian parameters (for line 3 (A1)) are given as minor and major axis ($\pm$0\farcs1) and position angle (PA, $\pm$4\degr).}
   \end{table}

\section{Discussion}

\subsection{Spectra and integrated intensity}
The different water lines toward source A show similar characteristics. The peak positions and line widths of the Gaussian fits to the 203~GHz \isowater\ and 225~GHz HDO lines are similar and agree with previous studies of other molecules toward IRAS~16293-2422 e.g., \citet{bisschop08,jorgensen11,jorgensen12,pineda12} (see Table~\ref{table:lines}). The mapped emission is compact and traces the warm water on scales $R$\textless200~AU. 

Toward source B, the 692~GHz \isowater\ line is detected in absorption. The spectral line is slightly red-shifted compared with $v_{LSR}$=2.7 \kms\ of source B, consistent with the picture of ongoing infall in this source \citep{jorgensen12,pineda12}. Integrating over a larger velocity interval reveals, in addition to the absorption, a blue-shifted emission peak, offset $\sim 0\farcs2$ south of the absorption (see Fig.~\ref{figure:spectra} and \ref{figure:moments}).

\subsection{Temperature and deuteration of water vapor}

To calculate the excitation temperature and column densities toward source~A we assume LTE conditions and that the emission at 203 and 225~GHz has the same extent as in the 692~GHz observations (see Table~\ref{table:lines}). Scaling the 203 GHz \isowater\ intensities to the 692 GHz source size gives an excitation temperature of $T_\mathrm{ex}=124\pm 12$~K for \isowater. If the 692~GHz line is indeed blended, it would imply that the intensity estimate, and also the excitation temperature, are upper limits. Halving the intensity of the 692~GHz line causes a drop to $T_\mathrm{ex}=107\pm 9$~K. The extent and peak location of the emission in source~A show that the warm water vapor is located at a small projected distance from the center of the protostar. The determined excitation temperature indicates that the distance is also small along the line of sight.

The estimated HDO column density is $4.8\times 10^{17}$~cm$^{-2}$, corrected for beam dilution when assuming that the extent of the emission is that of the 692~GHz \isowater\ observations and $T_\mathrm{ex}=124$~K. The gas-phase \normalwater\ column density is $5.3\times 10^{20}$~cm$^{-2}$ assuming the same as for HDO and that the isotopic abundance ratio of $^{16}$O/$^{18}$O is 560 in the local interstellar medium \citep{wilson94}. Assuming an uncertainty of about 20\% for the column densities, originating in the flux calibration, the best estimate of the HDO/\normalwater\ ratio is $(9.2\pm 2.6)\times 10^{-4}$.

Given the uncertainty in the determination of the excitation temperature, testing the effect of different temperatures is important. If the excitation temperature is as low as 80~K the HDO/\normalwater\ ratio becomes $7.8\pm 2.2\times 10^{-4}$. Increasing the excitation temperature to 300~K \citep{jorgensen12} increases the ratio to $1.1\pm 0.3\times 10^{-3}$. The conclusions in this paper do not change over this wide interval in $T_\mathrm{ex}$. In addition, since the 203~GHz \isowater\ and 225~GHz HDO observations have comparable beam and sources size, and arise from levels with similar energies, the deduced HDO/H$_2$O ratio is robust. However, if the HDO emission were more extended, its column density and the HDO/\normalwater\ ratio would be lower. In this scenario our inferred ratio is an upper limit.

In contrast with previous estimates based on models of single-dish observations \citep{parise05,coutens12}, the deduced HDO/\normalwater\ ratio in the warm gas of IRAS 16293-2422 is only slightly higher than found in Earth's oceans and by recent {\it Herschel} observations of comets. Given the possible systematic errors due to assumptions of the extent, they could be even closer. Within the statistical uncertainties our observed ratio for this protostar agrees with the earlier ratios for Oort cloud comets. Comparing these different ratios for water \emph{directly} assumes that the reservoirs, i.e., comets, Earth (planet) and inner protostellar region are linked and can be related.

The difference with the earlier estimates comes from the fact that those data are sensitive to much larger scales than the high-resolution interferometric observations presented here, which directly image the water emission on $25-280$~AU scales. Our interferometric observations provide a strong, model independent constraint on the deuteration of water in the innermost regions of protostars. That the low HDO/H$_2$O ratio in the warm gas is not much different from the cometary values is an indication that significant processing of the water between these early stages and the emerging solar system is not required.

Further high-resolution interferometric measurements toward larger samples of protostars will reveal whether the warm HDO/\normalwater\ ratio is similar in different protostars. In particular, future high angular resolution observations with ALMA will be able to resolve possible variations in the HDO/H$_2$O ratio with distance from the
central protostar and thereby show whether the slightly different ratios measured in different types of comets potentially could be related to their spatial origin in the protostellar envelope \citep{robert00}.

\begin{acknowledgements}
  We thank the referee, Al Wootten for insightful comments.
  The research at Centre for Star and Planet Formation is supported by
  the Danish National Research Foundation and the University of
  Copenhagen\rq{s} programme of excellence. This research was
  furthermore supported by a Junior Group Leader Fellowship from the
  Lundbeck Foundation to JKJ.  EvD acknowledges the Netherlands
  Organization for Scientific Research (NWO) grant 614.001.008 and EU
  FP7 grant 291141 CHEMPLAN.  This paper makes use of the following
  ALMA data: ADS/JAO.ALMA\#2011.0.00007.SV.  The authors are grateful
  to JAO CSV team for planning and executing the ALMA data, and to Tim
  van Kempen and Markus Schmalzl of the Allegro node for help with
  data reduction.  ALMA is a partnership of ESO (representing its
  member states), NSF (USA) and NINS (Japan), together with NRC
  (Canada) and NSC and ASIAA (Taiwan), in cooperation with the
  Republic of Chile. The Joint ALMA Observatory is operated by ESO,
  AUI/NRAO and NAOJ.  The Submillimeter Array is a joint project
  between the Smithsonian Astrophysical Observatory and the Academia
  Sinica Institute of Astronomy and Astrophysics and is funded by the
  Smithsonian Institution and the Academia Sinica.
\end{acknowledgements}

\bibliography{bibfile}

%

\Online

\begin{appendix}

\section{Tables}
   \begin{table*}
	\caption[]{Relevant parameters from the molecular line catalogs for the observed lines.}
         \label{table:observations}
         \begin{center}
         \begin{tabular}{lllllllll}
            \hline
            \noalign{\smallskip}
 Line		    	&   Species 	&	Frequency	&  Transition					& Line strength 	&  \eu	  	&	Beam 								& Resolution 	&  RMS	 			\\
 Id$^\mathrm{a}$	&		   	&	{[GHz]}		&  							& {[Debye$^2$]} & {[K]} 	&	Size (PA)							& {[\kms]} 	& {[$^\mathrm{b}$]} 	\\
            \noalign{\smallskip}
            \hline
            \noalign{\smallskip}
 1			 	&  \isowater & 203.40752   	& \trans{3}{1}{3}{2}{2}{0}  	& 0.34		& 203.7		& $3.1\arcsec\times5.0\arcsec$ (-25.3\degr)	& 0.30		& 208				\\
 2				& HDO 	    	& 225.89672   	& \trans{3}{1}{2}{2}{2}{1}	  	& 0.69		& 167.6	 	& $2.3\arcsec\times3.9\arcsec$ (15.0 \degr)	& 0.54 	 	& 190				\\
 3				& \isowater  	& 692.07914		& \trans{5}{3}{2}{4}{4}{1} 	& 1.26 		& 727.6 		& $0\farcs17\times0\farcs29$ (109.8\degr) 	& 0.40 		& 82					\\

            \noalign{\smallskip}
            \hline
         \end{tabular}
         \end{center}
         \begin{list}{}{}\setlength{\itemsep}{0.0ex}
	\item[$^{\mathrm{a}}$] Identifies the lines in Table~\ref{table:lines}.
	\item[$^{\mathrm{b}}$]mJy beam$^{-1}$ channel$^{-1}$
	\end{list}
   \end{table*}
   
\end{appendix}

\end{document}